\documentclass[12pt,preprint]{aastex}
\usepackage{natbib}
\usepackage{amssymb}
\usepackage{amsmath}
\usepackage{graphicx}

\def\aap{A\& A}
\def\aaps{A\& A Suppl.}
\def\apss{Astrophysics \& Space Science}

\def\araa{AnnRevA\& A}

\def\apj{ApJ}
\def\aj{AJ}

\begin{document}
\shorttitle{Explaining UXOR variability}
\shortauthors{Dullemond, v.d.~Ancker, Acke \& van Boekel}
\title{Explaining UXOR variability with self-shadowed disks}
\author{C.P.~Dullemond}
\affil{Max Planck Institut f\"ur Astrophysik, P.O.~Box 1317, D--85741 
Garching, Germany; e--mail: dullemon@mpa-garching.mpg.de}
\author{M.E.~van~den~Ancker}
\affil{European Southern Observatory, Karl-Schwarzschild Strasse 2, D-85748
Garching bei M\"unchen, Germany; e--mail: mvandena@eso.org}
\author{B.~Acke}
\affil{Instituut voor Sterrenkunde, KULeuven, Celestijnenlaan 200B, 
3001 Leuven, Belgium; e--mail: Bram.Acke@ster.kuleuven.ac.be}
\author{R.~van Boekel}
\affil{Sterrenkundig Instituut `Anton Pannekoek', Kruislaan 403,
NL-1098 SJ Amsterdam, The Netherlands; e--mail: vboekel@science.uva.nl}

\begin{abstract}
In this Letter we propose a new view on the phenomenon of
Algol-type minima in the light curves of UX Orionis stars. The idea is based
on the earlier proposal by various authors that UXORs are nearly-edge-on
disks in which hydrodynamic fluctuations could cause clumps of dust and gas
to cross the line of sight. However, early models of protoplanetary disks
were based on the notion that these have a flaring geometry. If so, then it
is mostly the outer regions of the disk that obscure the star. The time
scales for such obscuration events would be too long to match the observed
time scales of weeks to months. Recent 2-D self-consistent models of Herbig
Ae/Be protoplanetary disks (Dullemond \citeyear{dullemond:2002} henceforth
D02; Dullemond \& Dominik in prep., henceforth DD03), however, have
indicated that for Herbig Ae/Be star disks there exists, in addition to the
usual flared disks, also a new class of disks: disks that are fully
self-shadowed. For these disks only their puffed-up inner rim (at the dust
evaporation radius) is directly irradiated by the star, while the disk at
larger radius resides in the shadow of the rim. For these disks there exist
inclinations at which the line of sight towards the star skims the upper
parts of the puffed-up inner rim, while passing high over the surface of
outer disk regions. These outer disk regions therefore do not obscure the
star nor the inner disk regions, and small hydrodynamic fluctuations in the
puffed-up inner rim could cause the extinction events seen in UXORs. If this
idea is correct, it makes a prediction for the shape of the SEDs of
these stars. It was shown by D02/DD03 that flared disks have a strong far-IR
excess and can be classified as `Group I' (in the classification of Meeus et
al.~\citeyear{meeuswatersbouw:2001}), while self-shadowed disks have a
relatively weak far-IR excess and are classified as `Group II'. Our model
therefore predicts that UXORs belong to the `Group II' sources.  We show
that this correlation is indeed found within a sample of 86 Herbig Ae/Be
stars.
\end{abstract}

\section{Introduction}
UX Orionis objects (UXORs, see e.g.~reviews by Th\'e \citeyear{the:1994} and
Waters \& Waelkens \citeyear{waterswaelkens:1998}) are mostly intermediate
mass pre-main-sequence stars displaying a peculiar kind of photometric
variability: their V-band light curves are characterized by sudden drops in
brightness of up to 3 magnitudes with durations of days to many weeks.
These events are separated by relatively long periods of persistence. During
these so-called `Algol-type minima' (the name refers to the similarity of
the rapid decrease in brightness to those exhibited by eclipsing binaries
such as Algol) their spectrum is reddened, and is accompanied by a strong
increase in polarization (Grinin et al.~\citeyear{grininkiselev:1991}). In
very deep minima the star often becomes bluer again (e.g.~Bibo \& Th\'e
\citeyear{bibothe:1991}). The origin of these brightness drops and the
bluing effect has been debated for a long time. The currently favored view
is that variations of the column density of dust in the line of sight to the
star is to be held responsible. This idea was first put forward by Wenzel
(\citeyear{wenzel:1969}) and has since been worked out in more detail by
many authors (e.g.~Grinin \citeyear{grinin:1988}; Voshchinnikov
\citeyear{voshchinnikov:1990}; Grinin et al.~\citeyear{grininrost:1998}). If
these obscuration events are due to localized dust clumps of filaments
passing through the line of sight, then all phenomenological properties are
reproduced: the initial reddening due to dust absorption, and the increase
of polarization and later bluing due to the unobscured scattered
radiation. While this picture is attractive, it does not explain the nature
and origin of these dust clumps or filaments.

One of the leading theories is that Herbig Ae/Be stars are surrounded by
many large proto-cometary-clouds or cometary bodies (Grady et
al.~\citeyear{gradysitko:2000} and references therein). When one of these
objects happens to cross the line of sight towards the star, then an
absorption event is expected with precisely the characteristics seen in UXOR
stars. The absorption initially reddens the stellar light, but for very
large extinction the light scattered off other dust clouds starts to dominate,
restoring the color again. One of the main problems of this model is making
it consistent with the observed infrared SED. Most of these UXORs have
rather strong far-IR excess, with $L_{\mathrm{dust}}/L_{*}\simeq
0.2...0.3$. This ratio is an indication of the covering fraction of the dust
with respect to the star, so that for UXORs at least 20\% of the sky is
covered with dust grains. If this all is due to comets, their orbits must be
almost isotropically distributed, and there must be a very large number of
cometary bodies orbiting around the star. Moreover, the chemical composition
of gaseous material moving in and out of the line of sight does not appear
to resemble that expected in comets (Natta et al.~\citeyear{natgrintamb:2000}).

Currently a more favored model for the obscuring clouds/filaments seems to
be that of a protoplanetary disk seen nearly edge-on (Grinin et
al.~\citeyear{grininkiselev:1991}, \citeyear{grininrostop:1996}; Bertout
\citeyear{bertout:2000}; Natta \& Whitney \citeyear{nattawhitney:2000};
Kozlova et al.~\citeyear{kozlovagrinrost:2000}). If the inclination angle of
the disk is so large that the line of sight towards the star skims the
surface of the disk, then it is conceivable that hydrodynamic fluctuations
of the surface of the disk could cause dust filaments to pass through the
line of sight. Since protoplanetary disks are usually assumed to have a
flaring geometry (Kenyon \& Hartmann \citeyear{kenyonhart:1987}) dust
filaments held responsible for these obscuration events must reside in the
outer regions of the disk (Th\'e \citeyear{the:1994}).  A problem with this
picture is that one does not expect obscuration events of this kind to be on
a time scale of days or weeks. Turbulent eddies and filaments in the disk at
100 AU will tend to move across the star on a time scale of multiple
years. One would have to invoke turbulent eddies with a size of about 0.01
times the pressure scale height at 100 AU, and which are very compact
(hundreds of times the typical densities at those radii) in order to explain
these phenomena with the outer edge of the disk.

In this Letter we propose a new version of the nearly-edge-on disk
hypothesis, which does not suffer from the problems mentioned above, and
which in fact arises quite naturally from self-consistent models of 
protoplanetary disks around Herbig Ae/Be stars.

\section{The model}
\begin{figure*}
\centerline{
\includegraphics[width=8cm]{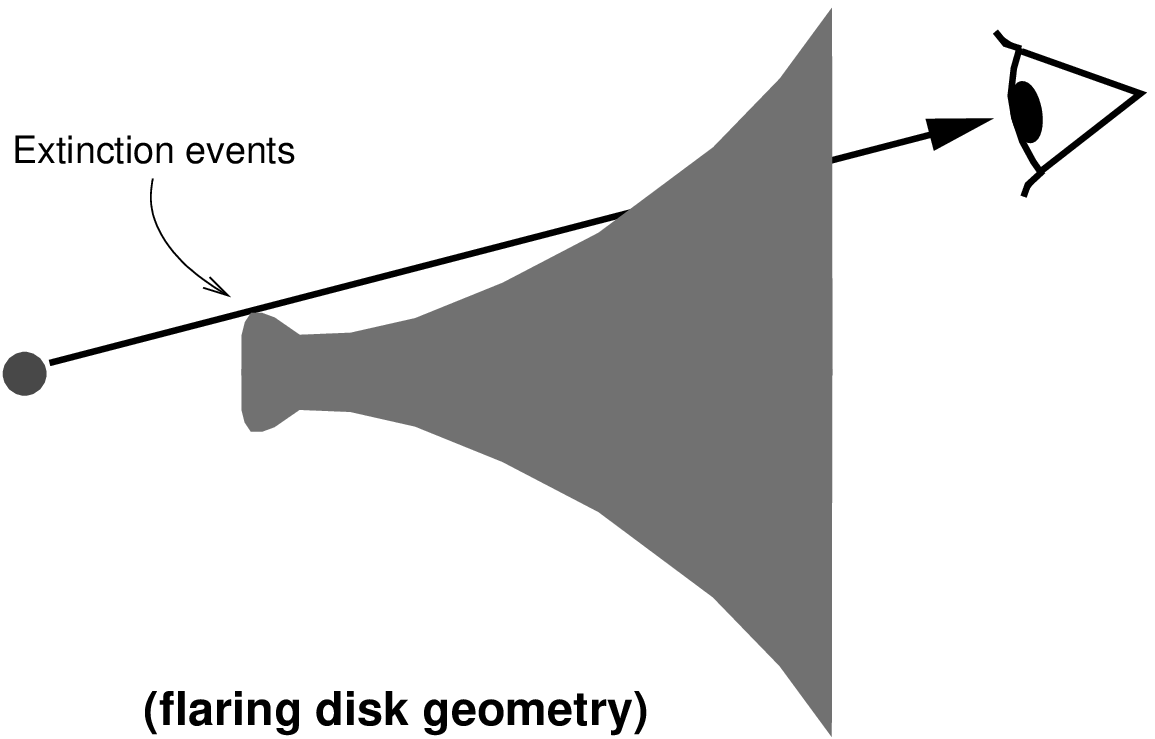}\hspace{4em}
\includegraphics[width=8cm]{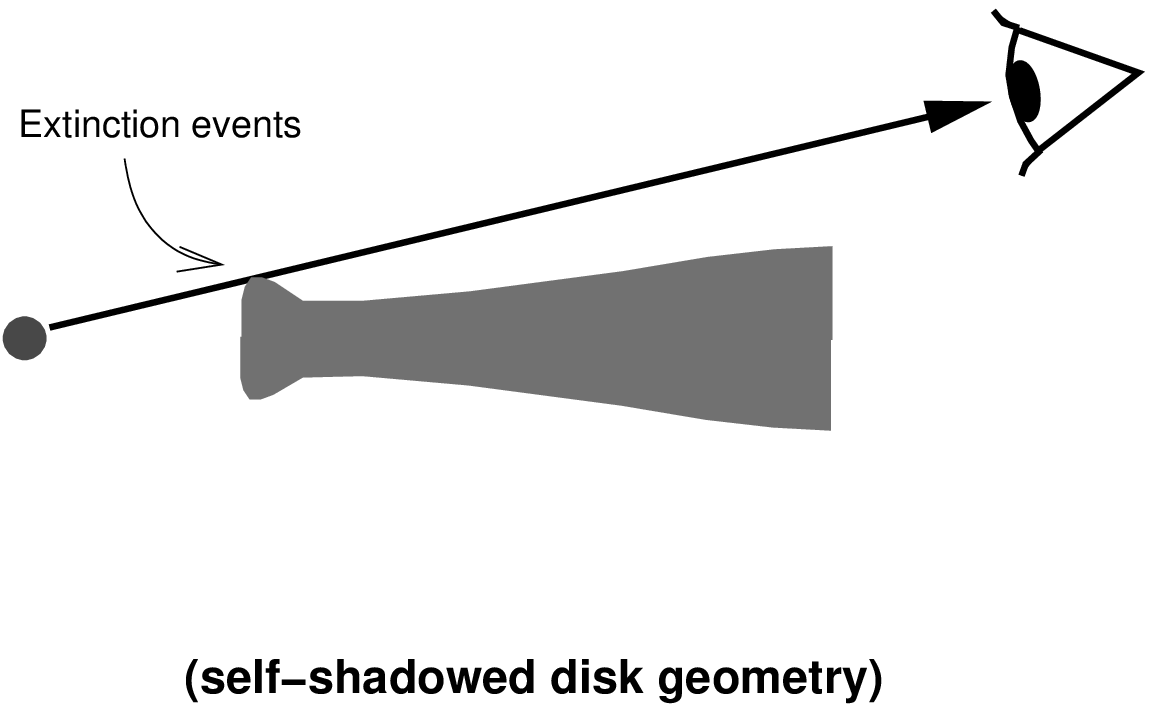}}
\caption{Pictographical illustration of the nearly-edge-on disk model. 
The left
panel shows that the puffed-up inner edge of the disk (a naturally arising
phenomenon in models of Herbig Ae/Be disks, see DDN01) may be responsible
for the UXOR extinction events, but that one needs inclination angles at
which the outer flaring part would have already completely occulted both the
central star and the inner rim. The right panel shows that the new disk
solutions found by D02/DD03 (the ``self-shadowed'' disks) do not have this
problem. Hydrodynamic fluctuations in the puffed-up inner rim can cause
short time scale extinction events, while the outer disk is not disturbing
the view anymore.
\label{fig-pictograms}}
\end{figure*}

The idea we wish to present in this Letter is inspired by recent models of
passive disks around Herbig Ae/Be stars (Dullemond, Dominik \& Natta
\citeyear{duldomnat:2001}, henceforth DDN01; Dullemond
\citeyear{dullemond:2002}). In these models, dust evaporation by the
radiation of the central star has removed the dust inward of the dust
evaporation radius. The dusty part of the disk therefore has an inner rim
(at around 0.5 to 1 AU from the central star) which is irradiated frontally
by the central star, and has a puffed-up geometry (Natta et
al.~\citeyear{nattaprusti:2001}; DDN01). This puffing-up is a result of the
fact that the rim is much hotter than the rest of the disk, since the latter
is only irradiated under a shallow angle (see e.g.~Calvet et
al.~\citeyear{calvetpatino:1991}; Chiang \& Goldreich
\citeyear{chianggold:1997}). Bell et
al.~(\citeyear{bellcassklhen:1997}) showed that viscous dissipation by
accretion could also cause the inner regions of the disk to puff up. The
puffed-up geometry of the hot inner rim makes it a good candidate for the
origin of the obscuring clouds. The time scales are right: a Kepler time
scale is a few months, so that it is very well possible that turbulent
filaments pass through the line of sight in a matter of days to weeks. Also
the densities are right: the inner rim is expected to have a very high
density, allowing that even relatively tenuous hydrodynamic perturbations
can have optical depths much larger than unity. The fact that this puffed-up
inner rim is a good candidate source for obscuring clouds was already
suggested by Natta et al.~(\citeyear{nattaprusti:2001}). But they admit that
for inclinations necessary for the inner rim to marginally obscure the star,
the outer flaring disk must already strongly obscure both the star and the
inner rim.  The star would then not have been classified as a Herbig Ae/Be
star.  A pictographical representation of the idea and the corresponding
problem with the flaring outer disk is shown in the left panel of
Fig.~\ref{fig-pictograms}.

\clearpage
\clearpage
A solution to this flaring disk problem may lie in recent 2-D models of
Herbig Ae/Be star disks (Dullemond \citeyear{dullemond:2002}, henceforth
D02; Dullemond \& Dominik in prep., henceforth DD03). In these models the
structure and SED of the disk is computed self-consistently by coupling 2-D
continuum radiative transfer to the equations of vertical hydrostatic
equilibrium. It was shown that, in addition to the flaring disk + inner rim
geometry, there can also exist disks that have sunk entirely into the shadow
of their own puffed-up inner rim. These ``self-shadowed disks'' have a
similar inner rim structure as the previous models, but they don't have the
flaring outer regions. For these disks one can therefore easily find
inclinations in which the line of sight skims the inner rim of the disk,
{\em without} passing through a flaring outer disk. In this case the idea,
that hydrodynamic turbulent filaments from the puffed-up inner rim 
are the root cause of UXOR variability, may in fact work. An illustration
is shown in the right panel of Fig.~\ref{fig-pictograms}.

Interestingly, this hypothesis makes a prediction for the shape of the SEDs
of UXORs. Self-shadowed disks were shown in D02/DD03 to have relatively weak
far-IR excess. They may be classified as ``Group II'' sources, in the
classification of Meeus et al.~(\citeyear{meeuswatersbouw:2001})
\footnote{Note that both Meeus et al.~{\em Group} I and {\em Group}
II sources are optically visible and hence belong to the
Lada~(\citeyear{lada:1987}) {\em Class} II YSOs.}. Flaring disks, on the other
hand, have a relatively strong far-IR excess, and belong to the ``Group I''
sources. The hypothesis put forward in this Letter therefore predicts that
UXORs belong to Group II. There may be marginal cases in which the flaring
is present, but weak enough to allow the observer to look through the
flaring part without too strong extinction. But these should be a
minority. The majority of UXOR sources should belong to Group II.

Incidentally, we note that emission from polyaromatic hydrocarbons (PAHs),
widely observed in Herbig Ae/Be stars, is believed to arise mainly in the
part of the surface layer of the disk that is exposed to direct stellar
radiation.  Therefore, our 2-D models predict that Group II sources should
have only weak or non-existent emission of PAHs, while Group I sources may
have strong emission of this kind (DD03). Our proposed model for UXOR
variability therefore also predicts UXORs to have weak to non-existent PAH
emission.

\section{SEDs of UXORs and other Herbig Ae/Be stars}
\label{sec-seds}

\clearpage
\begin{figure}
\centerline{
\includegraphics[width=6.5cm,angle=270]{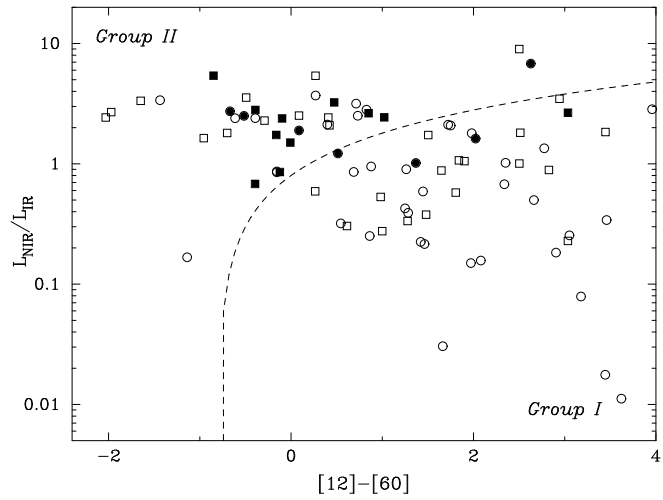}}
\caption{Distribution of the ratio of near-infrared to far-infrared 
luminosity versus {\em IRAS} [12]-[60] colour for 86 Herbig Ae/Be 
stars from the catalog of Th\'e et al. (1994).  Filled symbols
indicate UX Orionis stars.  The dashed line indicates the empirical 
separation between stars with a ``Group I'' energy distribution (i.e. 
relatively strong Far-IR excess) and the ``Group II'' sources. 
The squares are the sources for which we know that they are reasonably
isolated.
\label{fig-lnir-miras}}
\end{figure}
\clearpage
Guided by the hypothesis put forward in the previous section, we have
investigated the correlation between the shape of the infrared energy
distribution and the occurence of UXOR phenomena in HAeBes.  Following van
Boekel et al.~(\citeyear{vanboekelwaters:2003}), we characterize the
infrared energy distribution of HAeBes by two quantities: the ratio of
$L_{\rm NIR}$ (the integrated luminosity as derived from broad-band $J$,
$H$, $K$, $L$ and $M$ photometry from the literature) to $L_{\rm IR}$ (the
corresponding quantity derived from {\em IRAS} 12, 25 and 60~$\mu$m
photometry), and the (non color-corrected) {\em IRAS} [12]-[60] color.
Since Meeus et al.~Group I sources show an energy distribution closer to a
power-law than the more ``double-peaked'' energy-distributions of their
Group II sources, these two groups will naturally separate in a diagram of
$L_{\rm NIR}$/$L_{\rm IR}$ versus the {\em IRAS} [12]-[60] color. 
This will be discussed in detail in van Boekel et al.~in prep. In
Fig.~\ref{fig-lnir-miras} the diagram is shown for a sample of 86 HAeBes.

Using $JHKLM$ and {\em IRAS} photometry from literature, we have computed
$L_{\rm NIR}$/$L_{\rm IR}$ and {\em IRAS} [12]-[60] for all probable HAeBes
of spectral types B, A, and F from the catalogue of Herbig Ae/Be stars by
Th\'e et al.~(\citeyear{thedewinterperez:1994}) for which sufficient
photometry was available to compute these quantities.  Empirically, we find
that the line $L_{\rm NIR}$/$L_{\rm IR} > ([12]-[60]) + 0.9$ provides the
best separation between the sources known from visual inspection of their
energy distribution to belong to Group I and those belonging to Group II.

Using this relation, we find that our sample of 86 HAeBes contains 47 Group
I sources and 39 Group II sources.  If we define UX Orionis stars as stars
of spectral type B9 or later (earlier-type stars do not show the UXOR
phenomen; Bibo \& Th\'e 1991; van den Ancker et
al.~\citeyear{anckerwinterdjie:1998}) showing optical variations larger than
1 magnitude on time-scales of days to weeks, 18 of the sources in our sample
can be classified as UXOR.  Of those, 14 are located in the part of the
$L_{\rm NIR}$/$L_{\rm IR}$ versus {\em IRAS} [12]-[60] diagram occupied by
Meeus et al. Group II sources, whereas the other four are close to the line
separating groups I and II. If we limit ourselves to those sources for which
we know that they are reasonably `isolated' (i.e.~they are certain to be the
dominant IR source in the IRAS beam), then there are ten UXORs in Group II
and one in Group I.

\section{Discussion and conclusion}
On the basis of the findings of Section \ref{sec-seds} we conclude that the
2-D disk-model prediction, that UXOR-type phenomena should only occur in
self-shadowed disks, i.e.~the HAeBes with relatively modest
far-infrared excesses, is consistent with the observational data present in
literature.

We also note that a new study (Acke et al., in prep.) of all 48 Herbig Ae/Be
stars observed spectroscopically by the {\em Infrared Space Observatory}
finds that, while Group I sources often show strong PAH emission, these
spectral features are absent or weak in Group II sources. Moreover, they
find that the vast majority of the UXORs do not have strong PAH features in
their infrared spectra. This is consistent with the model predictions.

An important issue that we have not addressed so far is the percentage of
Group II sources that we expect on the basis of our model to display
UXOR-type variability. This question is difficult to answer, since this
requires knowledge of the turbulent behavior of the disk. Typically the
surface height of the inner rim of such a disk is of the order of
$H_{\mathrm{rim}}\sim 0.2 R_{\mathrm{rim}}$. If hydrodynamic fluctuations
are of the order of $\delta H_{\mathrm{rim}}\sim 0.1 R_{\mathrm{rim}}$, 
then one expects about 13\% of Group II stars to display UXOR variability. 
From Fig.~\ref{fig-lnir-miras} it seems that about 33\% of the Group II 
sources in the catalog of Th\'e et al. 1994) have UXOR variability.  
However, we note that the historical selection criteria for Herbig Ae/Be 
stars clearly favour strongly variable stars, such as UXORs.  Therefore 
it could well be that the real fraction of UXORs is smaller than the 
33\% in our current sample.  Clearly a study of the UXOR fraction in 
an unbiased sample of Herbig Ae/Be stars could place more stringent 
contraints on the magnitude of the hydrodynamic fluctuations required 
to explain the fraction of UXORs.

We stress that so far our model is not dependent on details of the
dust properties. However, for an exact prediction of the degree of
polarization and the detailed shape of color-magnitude diagrams, the
scattering properties of the dust grains, in addition to the geometry of the
disk, play an important role. In typical 2-D axisymmetric models of the
structure of disks of Herbig Ae stars described in D02/DD03 we find that the
disk covering fractions are within the range required by Natta \& Whitney
(\citeyear{nattawhitney:2000}) to reproduce the color magnitude diagrams and
polarization behavior of UXORs. Therefore we expect our model to be able
to reproduce this aspect of the observational characteristics of UXORs
as well.

Finally, for the model to work, the disk must have a puffed-up inner
rim. For Herbig Ae/Be stars this is a natural consequence of the 2-D
self-consistent model.  For T Tauri stars there might exist such a rim as
well, but it is far more difficult to find parameters for which the entire
disk lies in the shadow of the inner rim. On the basis of these arguments it
is therefore to be expected that T Tauri stars are only rarely seen to be
UXOR type stars.  There appear to be some indications that this is indeed
the case: Herbst et al.~(\citeyear{herbstherbst:1994}) report that the
dominant sources of photometric variations seen in T Tauri stars are
rotational modulation due to cool spots on the stellar surface, and changes
in the excess or veiling continuum.  UXOR variations, although present in
their sample, appear to be rare, and are limited to stars with spectral type
earlier than K0.

We conclude that the explanation for UXOR variability presented in this
paper seems to work better than explanations published in the literature so
far. It is a natural consequence of self-consistent 2-D models of disks
around Herbig Ae/Be stars, and it makes a number of predictions, 
which seem to be confirmed by observations.

\begin{acknowledgements}
CPD wishes to thank A.~Natta and C.~Dominik for discussions and suggestions.
\end{acknowledgements}

\end{document}